\documentclass[hyper,letterpaper,12pt]{article}
\usepackage{a4wide}
\usepackage{amsmath}
\usepackage[
      colorlinks=true,
      linkcolor=blue,
      urlcolor=blue,
      filecolor=black,
      citecolor=blue,
            ]{hyperref}
\usepackage{color}
\usepackage{graphicx}
\usepackage{amsfonts}
\usepackage{amssymb}
\usepackage{epstopdf}

\def\beq{\begin{equation}}
\def\eeq{\end{equation}}

\begin{document}
\title{\bf \Large Thermodynamics of Conformal Anomaly Corrected Black Holes in AdS Space}

\author{\large
~Rong-Gen Cai\footnote{E-mail: cairg@itp.ac.cn}
\\
\small  State Key Laboratory of Theoretical Physics,\\
\small Institute of Theoretical Physics, Chinese Academy of Sciences,\\
\small Beijing 100190,  China \\
\small  King Abdulaziz University, Jeddah, Saudi Arabia }
\maketitle

\begin{abstract}
\normalsize We present exact analytical black hole solutions with conformal anomaly in AdS space and discuss the thermodynamical properties  of
these black hole solutions. These black holes can have a positive, zero and negative constant curvature horizon, respectively.
For the black hole with a positive constant curvature horizon, there exists a minimal horizon determined by the coefficient of the
trace anomaly, the black hole with a smaller horizon is thermodynamically unstable, while it is stable for the case with a
larger horizon. The Hawking-Page transition happens in this case. For the black hole with a Ricci flat horizon, the black hole
is always thermodynamically stable and there is no Hawking-Page transition. In the case of the black hole with a negative constant
curvature horizon, there exists a critical value for the coefficient of the trace anomaly, under this critical value, the black hole
is always thermodynamical stable and the Hawking-Page transition does not happen. When the coefficient is beyond the critical value,
the black hole with a smaller horizon is thermodynamically unstable, but it becomes stable for the case with a
larger horizon, the Hawking-Page transition always happens in this case. The latter is a new feature for the black holes with a negative constant curvature horizon.
\end{abstract}

\newpage
\section{Introduction}

The AdS/CFT correspondence~\cite{Maldacena:1997re,Gubser:1998bc,Witten:1998qj} conjectures that string/M theory in an anti-de Sitter (AdS) space (times some compact space) is dual to a strongly coupled  conformal field theory (CFT) living on the AdS boundary. In this sense, thermodynamics of black holes in AdS space can be identified with the one of the dual strongly coupled CFTs~\cite{Witten}, while the latter is not able to be obtained through conventional perturbative  quantum field theory.  This is one of reasons for many recent studies of black holes in AdS space.  On the other hand, it is also of great interest to investigate black holes in AdS space in its own right. Compared
to its counterpart in asymptotically flat spacetime, the horizon of black holes in AdS space has much rich topological structure: AdS black holes can have positive, zero and negative constant curvature horizon. These black holes have been discussed extensively, for example,  in Einstein-(Maxwell) theory~\cite{Lemos1}-\cite{Birm}, Einstein-Maxwell-Dilaton theory~\cite{CZ,CJK}, Wely conformal gravity~\cite{Klemm}, Gauss-Bonnet gravity~\cite{BD,Cai1,Nojiri,Cho}, and more general Lovelook gravity~\cite{BTZ,Cai2,CS,ATZ}, etc.

It turns out that thermodynamical properties of AdS black holes crucially depend on horizon structure. For (positive constant curvature) spherical horizon black holes in AdS space, there exists a minimal black hole temperature, under which there is no black hole solution.
The black hole is thermodynamically unstable for smaller black hole horizon, while it is thermodynamically stable for larger horizon. Therefore there exists a so-called Hawking-Page phase transition~\cite{HP} between the thermal gas and stable large black hole in AdS space. This phase transition can be identified with the confined/deconfined transition of gauge field in the dual CFT side~\cite{Witten}.
For the black holes with (zero curvature) Ricci flat horizon and (negative curvature) hyperbolic horizons, however, they are always thermodynamically stable (see, for example, \cite{Birm}). This means that the black hole phase is always dominated in the dual conformal field theory side. In this case, the Hawking-Page phase transition will not happen.

Conformal anomaly~\cite{Duff} plays an important role in quantum field theories in curved spacetime and has various applications in black hole physics, cosmology, string theory, and statistical mechanics, etc.
 The effective action associated with the trace anomaly is non-local, but in four dimensional spacetime case, it can be written in a local form by introducing two auxiliary scalar fields (see \cite{EM1,EM2} and reference therein). The quantum trace anomaly can give rise to
 macroscopic effects. To study the back reaction of the trace anomaly, one has to solve the equations of motion for these scalar fields in some curved or topological nontrivial backgrounds. It turns out that at the horizon, the quantum fluctuations of the scalar degrees of freedom world be large~\cite{EM2,AMV}.  These two scalar fields are determined by boundary conditions, which brings back the non-local
 effect of the trace anomaly. Furthermore, the equations of motion from the effective action of the trace anomaly is quite complicated, therefore in general it is impossible to obtain exact analytical solutions of Einstein equations with
 the trace anomaly.  In \cite{CCO}, with an additional assumption the authors are able to find an exact analytical
black hole solution to the Einstein equations with the trace anomaly. There some interesting properties of the trance anomaly corrected black hole are observed.  Some further discussions can be found, for example, in \cite{AEM}-\cite{ADM}.

In this note we are going to generalize the black hole solution presented in \cite{CCO} to the case with a negative cosmological constant
and to study the thermodynamical properties of the trace anomaly corrected black holes in AdS space. This note is organized as follows.
To be self-contained, in section~\ref{sect2} we will briefly review the procedure to solve the Einstein equations with the trace anomaly and present the
exact analytical black hole solutions. In section~\ref{sect3} we study the thermodynamical properties of these black holes with  positive,
zero and negative constant curvature horizon, respectively.  There we will see that the trace anomaly changes qualitatively thermodynamical
properties of the black holes, compared to the case without the trace anomaly. Section~\ref{sect4} is devoted to the conclusions and further
discussions.


\section{Black Hole Solutions}
\label{sect2}
In four dimensional spacetime, one loop quantum correction leads to a trace anomaly of the stress-energy tensor of conformal field
theory. The trance anomaly has the form~\cite{Duff,DS}
\begin{equation}
\label{eq1}
g^{\mu\nu}\langle T_{\mu\nu}\rangle=  \beta I_4-\alpha E_4,
\end{equation}
where $I_4=C_{\mu\nu\lambda\sigma}C^{\mu\nu\lambda\sigma}$ and $E_4=R^2-4 R_{\mu\nu}R^{\mu\nu}+R_{\mu\nu\lambda\sigma}R^{\mu\nu\lambda\sigma}$, $\beta$ and $\alpha$ are two positive constants related to the content of the conformal field theory:
\begin{equation}
\beta= \frac{1}{120(4\pi)^2}(n_0 +6n_{1/2}+12n_1), \ \ \ \alpha= \frac{1}{360 (4\pi)^2}(n_0+11n_{1/2} +62n_1),
\end{equation}
where $n_0$, $n_{1/2}$ and $n_1$ are the number of scalar, Dirac fermion and vector fields, respectively, in the conformal field theory. For example, for the ${\cal N}=4$ $SU(N)$ Supersymmetric Yang-Mills  theory, one has $n_0=6N^2$, $n_{1/2}=2N^2$ and $n_1=N^2$ in the large $N$ limit.
The first term in (\ref{eq1}) is called type B anomaly, while the second one is called type A anomaly~\cite{DS}.

Having considered the trace anomaly, the Einstein equations should be written as
\begin{equation}
\label{eq3}
R_{\mu\nu}-\frac{1}{2}Rg_{\mu\nu}+\Lambda g_{\mu\nu} =8\pi G \langle T_{\mu\nu}\rangle,
\end{equation}
where $G$ is the Newton gravitational constant, $\Lambda$ is the cosmological constant, and $ \langle T_{\mu\nu}\rangle$ is the corresponding
stress energy tensor associated with the trace anomaly.
In this paper we are interested in the case with a negative cosmological constant, $\Lambda=-3/l^2$. With only the trace (\ref{eq1}), in general it is impossible to determine all components of the stress-energy tensor associated with the trace anomaly, in the case without introducing two auxiliary scalar fields~\cite{EM1,EM2}. In \cite{CCO}, the authors are able to obtain an exact analytical solution to the equations (\ref{eq3}) in the static spherically symmetric spacetime with two additional assumptions: (i) The stress-energy tensor is covariant conserved, namely, $\nabla_{\mu}\langle T^{\mu\nu}\rangle =0$; (ii) The time-time component of the stress-energy tensor and the radial-radial component are equal in the spacetime, that is, $\langle T^t_{\ t}(r)\rangle =\langle T^r_{\ r}(r)\rangle $. The assumption (i) manifestly holds because the left hand side of the equations are of the same property,  while the assumption (ii)
 is imposed in order to find an exact analytical solution of the Einstein equations. Here we would like to stress that introducing the
 assumption (ii) is just for easily solving the equations, it is not a physical requirement. We now generalize the solution presented in \cite{CCO} into the case with a cosmological constant.

Consider the static spherically symmetric spacetime of the form
\begin{equation}
\label{eq4}
ds^2 = -f(r) dt^2 +\frac{1}{g(r)}dr^2 +r^2 d\Omega_{2k}^2,
\end{equation}
where $f(r)$ and $g(r)$ are two functions of the radial coordinate $r$, and $d\Omega_{2k}^2$ is the line element
of a two-dimensional Einstein constant curvature space with scalar curvature $2k$. Without loss of generality, one may
take $k$ to be $1$, $0$ and $-1$, respectively. In that case, they correspond to a sphere,  Ricci flat space, and negative curvature space, respectively. According to the symmetry of metric (\ref{eq4}), we define
\begin{equation}
\label{eq5}
\langle T^{t}_{\ t}\rangle =-\rho(r),\ \ \langle T^r_{\ r}\rangle= p(r),\ \  \langle T^{\theta}_{\ \theta}\rangle=\langle T^{\phi}_{\ \phi}\rangle =p_{\bot}(r),
\end{equation}
and other components must vanish. Under the assumption (ii), one can easily show that $f(r)=g(r)$ (see, for example, \cite{CJS}). Furthermore, combining
the covariant conservation of the stress-energy tensor with the trace anomaly (\ref{eq1}), we have
\begin{eqnarray}
\label{eq6}
&& -\rho+p +2 p_{\bot}=-\frac{2\alpha}{r^2} ((1-f)^2)'', \nonumber\\
&& 4 f (p-p_{\bot}) +(\rho +p)r f'+2 r f p'=0,
\end{eqnarray}
where we have set $\beta=0$ and the prime stands for the derivative with respect to the radial coordinate $r$.  With the assumption (ii), $\rho=-p$, we can have from (\ref{eq6}) that
\begin{equation}
rp' +4 p +\frac{2 \alpha}{r^2}((1-f)^2)''=0.
\end{equation}
This equation can be easily solved with the solution
\begin{equation}
\label{eq8}
p= \frac{2\alpha}{r^4} (1-f) (1-f+2r f')-\frac{q^2}{r^4},
\end{equation}
where $q^2$ is an integration constant. Thus we are able to obtain the other nonvanishing components of the stress-energy tensor associated with the trance anomaly as
\begin{eqnarray}
\label{eq9}
&& \rho =-p, \nonumber \\
&& p_{\bot}= -2 p -\frac{\alpha}{r^2}((1-f)^2)''.
\end{eqnarray}
Substituting (\ref{eq8}) and (\ref{eq9}) into (\ref{eq3}), we can obtain the solution to the Einstein equations (\ref{eq3}) as
\begin{equation}
\label{eq10}
f(r)=g(r)=k-\frac{r^2}{4\tilde\alpha}\left(1\pm \sqrt{1+\frac{8\tilde \alpha}{l^2}-\frac{16\tilde{\alpha}GM}{r^3}+\frac{8\tilde{\alpha}Q^2}{r^4}}\right),
\end{equation}
where $M$ is an integration constant, $\tilde {\alpha}= 8\pi G \alpha$ and $Q^2=8\pi Gq^2$. The solution has two branches with signs ``$\pm$" in (\ref{eq10}). When $M=Q=0$, the solution reduces to
\begin{equation}
f(r) = k-\frac{r^2}{4\tilde \alpha}\left (1 \pm \sqrt{1 +\frac{8\tilde \alpha}{l^2}}\right).
\end{equation}
In this case, the branch with sign ``$-$" corresponds to an AdS spacetime with the effective radius $l_{\rm eff}=\sqrt{4 \tilde {\alpha}/(\sqrt{1+8\tilde{\alpha}/l^2}-1)}$, while the branch with sign ``$+$" a de Sitter spacetime with effective radius $l_{\rm eff}= \sqrt{4 \tilde {\alpha}/(1+\sqrt{1+8\tilde{\alpha}/l^2})}$. Similar case happens in the Gauss-Bonnet gravity~\cite{Cai1}. Therefore we suspect that the solution with sign ``$+$" in (\ref{eq10}) might be unstable. Thus in the following we consider the case with sign ``$-$". In the limit $\tilde \alpha \to 0$, the solution reduces to
\begin{equation}
f(r)= k+\frac{r^2}{l^2} -\frac{2GM}{r}+\frac{Q^2}{r^2}.
\end{equation}
Clearly it is nothing but the Reissner-Nordstr\"om-AdS black hole solution. This suggests that the integration constant $M$ in (\ref{eq10}) is
just the mass of the solution. An interesting point is that when $\alpha=0$, one should get the Schwarzschild-AdS solution for the
Einstein equations (\ref{eq3}); this seemingly indicates that we should take the other integration constant $Q=0$ in the solution (\ref{eq10}). As pointed out in \cite{CCO}, however, this is not true. The reason is as follows. The term associated with the
constant $Q$ is a
``dark radiation" one with energy density and pressure as $\rho_d=p_{\bot}=-p= q^2/r^4$. Such a term obeys the assumptions (i) and (ii), and is consistent with the symmetry of the metric (\ref{eq4}). Therefore one is not able to exclude such a term in the solution in
the present setup. To more clearly see the meaning of the term, one can show that if a Maxwell field is present, the electric charge square
$Q_e^2$ for a static electric field will appear in the same place as $Q^2$ in the solution (\ref{eq10})~\cite{CCO}. Therefore the integration constant $Q$ can be interpreted as a U(1) conserved charge of the conformal field theory which leads to the trace anomaly.
 However, in what follows, we will not consider the effect of
the integration constant $Q$ since it is undistinguished from the electric charge $Q$ of the electric field. That is to say, we will set $Q=0$ in the following discussions.

\section{Thermodynamics of the Black Holes}
\label{sect3}

In this section we will study the effect of the trace anomaly on the thermodynamics of the AdS black holes. Suppose the solution (\ref{eq10}) describes a black hole with horizon $r_+$ and it is asymptotically AdS. One then has $f(r_+)=0$ and $f(r>r_+) >0$.  The black hole mass $M$ can be expressed in terms of the horizon radius as
\begin{equation}
\label{eq13}
M= \frac{r_+}{2G}\left( k +\frac{r_+^2}{l^2}-\frac{2 \tilde {\alpha}k^2}{r_+^2}\right).
\end{equation}
When $\tilde \alpha=0$, it reduces to the one for the Schwarzschild-AdS black hole. The Hawking temperature of the black hole can be easily
calculated with the result
\begin{equation}
\label{eq14}
T=  \frac{r_+}{4\pi (r_+^2-4\tilde{\alpha}k)}\left( k+\frac{3r_+^2}{l^2} +\frac{2\tilde{\alpha}k^2}{r_+^2}\right).
\end{equation}
In Einstein's general relativity, the Bekenstein-Hawking entropy of a black hole satisfies the so-called area formula: it is given by
one quarter of the horizon area. In the present case, due to the presence of the trace anomaly, we do not expect that the entropy of the
black hole still obeys the area formula. Instead, the first law of black hole thermodynamics, $dM=TdS +\cdots$, is always satisfied
for all black holes, here $\cdots$ stands for some work terms, if any.  Considering this fact and integrating the first law for our black hole solution, we obtain
\begin{equation}
\label{eq15}
S= \int T^{-1}\left(\frac{\partial M}{\partial r_+}\right) dr_+ = \frac{\pi r_+^2}{G}-\frac{4\pi \tilde {\alpha}k}{G} \ln r_+^2 +S_0,
\end{equation}
where $S_0$ is an integration constant. Here we cannot fix the constant $S_0$ by physical consideration because of the existence of
the logarithmic term. The black hole entropy can also be expressed in terms of the horizon area $A=4\pi r_+^2$ as
\begin{equation}
\label{eq16}
S= \frac{A}{4G}-\frac{4\pi \tilde{\alpha}k}{G}\ln \frac{A}{A_0},
\end{equation}
where $A_0$ is a constant with dimension of $[\rm length]^2$.

The local stability of black hole thermodynamics is determined by heat capacity of the black hole. In our case, the heat capacity of the
black hole is given by
\begin{equation}
\label{eq17}
C= T \left(\frac{\partial S}{\partial T}\right) = \frac{8\pi^2 T (r_+^2-4\tilde{\alpha}k)^3}{ G r_+^3}\left( -k -\frac{10 \tilde{\alpha}k^2}{r_+^2} +
\frac{3r_+^2}{l^2}-\frac{36 \tilde{\alpha}k }{l^2} +\frac{8\tilde{\alpha}^2k^3}{r_+^4}\right)^{-1}.
\end{equation}
And the global stability of black hole thermodynamics can be checked by studying its free energy, defined as
\begin{equation}
F=M-TS.
\end{equation}
Below we will discuss thermodynamical properties of the black hole solutions with $k=1$, $0$ and $-1$, respectively.

\subsection{The case of $k=1$}

In this case, in order to have a positive Hawking temperature, we see from (\ref{eq14}) that the black hole solution has a minimal
horizon radius $r_{\rm min}= 2 \sqrt{\tilde{\alpha}}$. As a consequence, the black hole has the minimal mass as
 \begin{equation}
 M_{\rm min}=\frac{\sqrt{\tilde{\alpha}}}{2G}\left(1+\frac{8\tilde{\alpha}}{l^2}\right).
 \end{equation}
 When $l^2 \to \infty$, the minimal mass becomes $\sqrt{\tilde{\alpha}}/2G$.
 For the minimal black hole, its Hawking temperature diverges. The existence of the minimal horizon radius implies
 that the black hole solution (\ref{eq10}) has a mass gap: when $M=0$, the solution is an AdS space with effective radius $l_{\rm eff}=\sqrt{4 \tilde {\alpha}/(\sqrt{1+8\tilde{\alpha}/l^2}-1)}$; when $M \ge M_{\rm min}$, the solution describes an asymptotically AdS black hole;
 and when $ 0<M<M_{\rm min}$, the solution describes a naked singularity. Clearly the new feature of the existence of mass gap
  arises totally
 due to the trace anomaly.  In addition, let us notice that besides the singularity at $r=0$,
 the solution (\ref{eq10}) has the other singularity at $r_s$, where the square root in (\ref{eq10}) vanishes:
 \begin{equation}
 r_s^3 =\frac{8\tilde{\alpha}r_+}{1+8\tilde{\alpha}/l^2}\left(k+\frac{r_+^2}{l^2}-\frac{2\tilde{\alpha}k^2}{r_+^2}\right).
 \end{equation}
 One can see from (\ref{eq10}) that the singularity is always covered by the black hole horizon with radius $r_+$, when the latter exists.

The thermodynamic behavior of the conformal anomaly corrected black hole is qualitatively the same as the one for the Schwarzschild-AdS black hole, only the difference is that there does not exist the minimal horizon in the latter case.  To see this, we plot the temperature of the black hole in Fig.~\ref{T_{k=1}}. Note that the case with $\tilde {\alpha}=0$ corresponds to the Schwarzschild-AdS black hole. We can see from the figure that the temperature goes to infinity as $r_+ \to 2\sqrt{\tilde{\alpha}}$ for small black holes with $r_+ \ll l$, while for large black holes with $r_+ \gg l$, the temperature diverges linearly with respect to $r_+$.
In addition, like the case of the Schwarzschild-AdS black hole, there always exists a minimal temperature for the conformal anomaly corrected black hole. Under this minimal temperature, there does not exist any black hole solution.
\begin{figure}[h]
\centering
\includegraphics[scale=0.92]{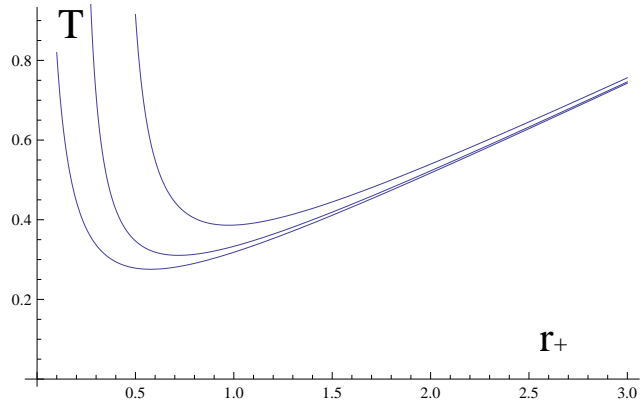}
 \caption{\label{T_{k=1}} The Hawking temperature of the black hole with $k=1$. The curves from up to down correspond to the cases
 with $\tilde {\alpha}/l^2=0.1$, $0.01$ and $0$, respectively. Here the  horizon radius is scaled by $1/l$, while the temperature is scaled by $l$. }
\end{figure}

The minimal temperature is determined by the divergence point of the heat capacity (\ref{eq17}), where the denominator in (\ref{eq17})
vanishes:
\begin{equation}
-1 -\frac{10 \tilde{\alpha}}{r_m^2} +
\frac{3r_m^2}{l^2}-\frac{36 \tilde{\alpha} }{l^2} +\frac{8\tilde{\alpha}^2}{r_m^4}=0.
\end{equation}
The above equation has a positive real root, which gives the black hole horizon corresponding to the minimal temperature.  When $\tilde{\alpha}=0$,  the solution is $r_m=l/\sqrt{3}$. Clearly we can see from the figure that when $r_{\rm min} <r_+ <r_m$, the
black hole has a negative heat capacity, indicating the local instability of the black hole, while the black hole is local
stable with a positive heat capacity as $r_+>r_m$.

Now we turn to the global thermodynamical stability of the black hole by studying its free energy $F=M-TS$.
Our aim is to see whether the presence
the logarithmic term in the black hole entropy modifies qualitatively the thermodynamical behavior of the Schwarzschild-AdS black hole. Unfortunately due to the presence of the integration constant $A_0$, we are not able to uniquely specify the free energy. In Fig.~\ref{F_{k=1}} we plot the free
energy in the case of $\tilde{\alpha}/l^2=0.01$ (corresponding minimal horizon radius $r_{\rm min}=0.2 l$), with three different choices of the constant $A_0/l^2=0.01$, $0.5$ and $100$, respectively.
While the behavior of the free energy is qualitatively similar to the case of $\tilde{\alpha}/l^2=0.01$ for other values of $\tilde{\alpha}$, it crucially depends on the constant $A_0$: although the free energy is always negative for larger black holes,  its behavior is qualitatively different in the region of smaller black holes.

\begin{figure}[h]
\centering
\includegraphics[scale=0.92]{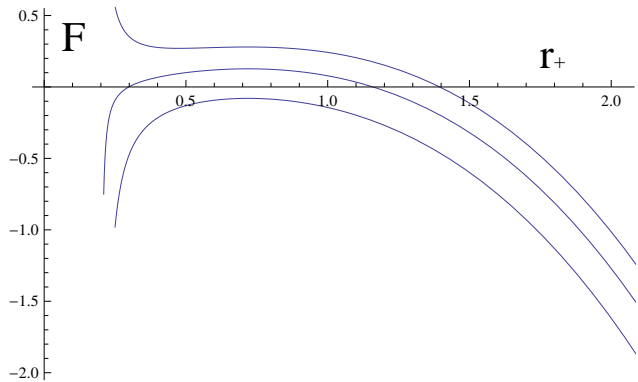}
 \caption{\label{F_{k=1}} The free energy of the black hole with $k=1$ and $\tilde{\alpha}/l^2=0.01$. The curves from up to down correspond to the cases
 with $A_0/l^2=0.01$, $0.5$ and $100$, respectively. Here the horizon radius is scaled by $1/l$, while the free energy is scaled by $G/l$. }
\end{figure}

For a smaller $A_0$ (for example, $A_0/l^2=0.01$ in Fig.~\ref{F_{k=1}}), the free energy behaves like the one of the Schwarzschild-AdS black hole: for smaller black holes, the free energy is positive, which indicates the global thermodynamical instability of the black hole, while it is negative for larger black holes, and the Hawking-Page phase transition happens when the free energy changes its sign. For a larger $A_0$ (for example, $A_0/l^2=100$ in Fig.~\ref{F_{k=1}}), the free energy is always negative. In that case, the Hawking-Page phase transition will not appear. The case with a middle $A_0$ (for example, $A_0/l^2=0.5$ in Fig.~\ref{F_{k=1}}) is very interesting: the free energy is negative for larger black holes and smaller black holes, but it is positive between them. In that case, there seemingly exist two Hawking-Page phase
transitions because the free energy changes its sign twice. At the moment we are not able to figure out which case discussed above is more
physical since one cannot specify the constant $A_0$.  But we will have some to say on this issue in the last section of the paper.

\subsection{The case of $k=0$}

This case is much simple than the other two cases. In this case, the Hawking temperature and entropy of the black hole are simply given by
\begin{equation}
T= \frac{3r_+}{4\pi l^2}, \ \ \ S= \frac{A}{4G},
\end{equation}
Note that in this case the logarithmic term disappears in the black hole entropy and here we have fixed the integration constant $S_0=0$ by requiring that the black hole entropy vanishes as the horizon $r_+ \to 0$. The temperature and entropy  of the black hole
are exactly the same as
those of the  Schwarzschild-AdS black hole with a Ricci flat horizon. The horizon radius $r_+$ has a relation with the black hole mass $M$ as $r_+=(2GMl^2)^{1/3}$. The heat capacity is always positive while the free energy $F=-r_+^3/(4Gl^2)$ is always negative. The Hawking-Page phase transition does not appear in this case. This indicates
that in this case the conformal anomaly corrected black hole is not only local stable, but also global stable thermodynamically, and the black hole phase is always dominated in the dual conformal field side.

\subsection{The case of $k=-1$}
This case is very interesting as we will see. In this case, the Hawking temperature is given by
\begin{equation}
\label{eq22}
T= \frac{r_+}{4\pi (r_+^2+4 \tilde{\alpha})}\left(-1+\frac{3r_+^2}{l^2}+\frac{2 \tilde{\alpha}}{r_+^2}\right).
\end{equation}
The behavior of the temperature heavily depends on the parameter $\tilde \alpha$. Let us first notice that for $T=0$, the equation (\ref{eq22}) has two solutions
\begin{equation}
\label{eq23}
r^2_{1,2}=\frac{l^2}{6} \left( 1 \pm \sqrt{1-\frac{24\tilde{\alpha}}{l^2}}\right).
\end{equation}
When $\tilde{\alpha}/l^2<1/24$,  we have two positive real roots (we denote the larger one by $r_2$ and smaller one by $r_1$); when $\tilde{\alpha}/l^2=1/24$, we have two degenerated real roots; while as $\tilde{\alpha}/l^2>1/24$, we have no real roots. Considering the fact that black hole temperature is always required to be positive,  the above analysis
tells us that when $\tilde{\alpha}/l^2<1/24$, the black hole has a minimal horizon given by $r_2$, for which the black hole has a vanishing
temperature. Clearly when $\tilde{\alpha}/l^2=1/24$, the black hole with horizon $r_2=l/\sqrt{6}$ has also a vanishing temperature. On the other hand, as $\tilde{\alpha}/l^2>1/24$, there does not exist any minimal horizon, the Hawking temperature is always positive. We plot in Fig.~\ref{T_{k=-1}} the behavior of the temperature with respect to the horizon radius $r_+$ for three typical parameters $\tilde{\alpha}/l^2=0.1$, $1/24$ and $0.02$.

\begin{figure}[h]
\centering
\includegraphics[scale=0.82]{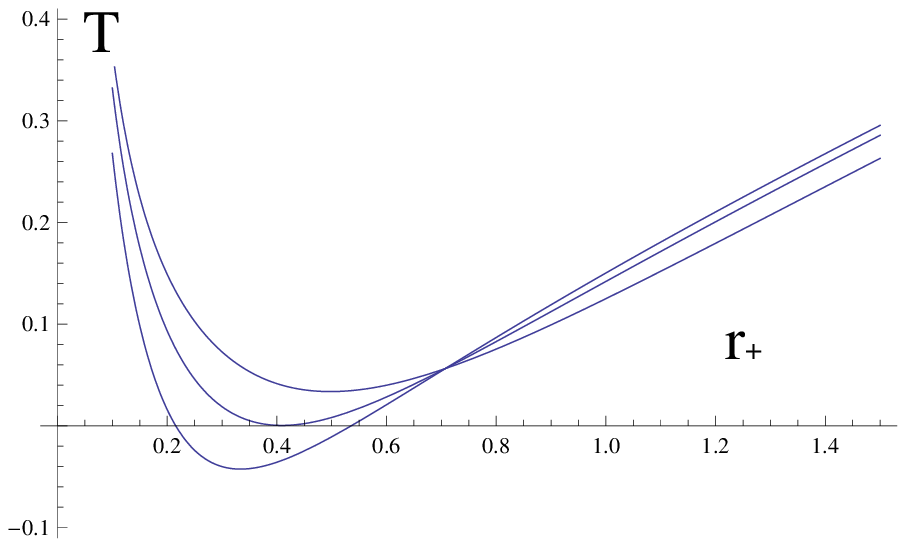}
\caption{\label{T_{k=-1}}The Hawking temperature of the black holes with $k=-1$. The three curves from up to down in the left side correspond to the cases with $\tilde {\alpha}/l^2=0.1$, $1/24$ and $0.02$, respectively. Here the  horizon radius is scaled by $1/l$, while the temperature is scaled by $l$.}
\end{figure}

(1) When $\tilde{\alpha}/l^2<1/24$, we can see from the figure that the temperature starts from zero at the minimal horizon $r_2$, which corresponds to an extremal black hole,  and increases monotonically as $r_+$ grows. This means that the black hole is always local stable thermodynamically.  (2) When $\tilde{\alpha}/l^2=1/24$, the temperature decreases monotonically from infinity at $r_+=0$, till to
zero at $r_2=l/\sqrt{6}$, and then increases monotonically when $r_+$ increases. This behavior implies that when $r_+<r_2$, the black hole
is local unstable with a negative heat capacity, while it is local stable as $r_+>r_2$. At $r_+=r_2$, the heat capacity diverges. (3) When $\tilde{\alpha}/l^2>1/24$, there exists a minimal temperature at $r_+=r_m$ satisfying
\begin{equation}
\label{eq24}
1 -\frac{10 \tilde{\alpha}}{r_m^2} +
\frac{3r_m^2}{l^2}+\frac{36 \tilde{\alpha} }{l^2} -\frac{8\tilde{\alpha}^2}{r_m^4}=0.
\end{equation}
Under this minimal temperature there does not exist any black hole solution. From Fig.~\ref{T_{k=-1}} one can easily see that
when $r_+<r_m$, the black hole has a negative heat capacity and is local unstable, while it is local stable with a positive heat capacity
as $r_+>r_m$. We find that the temperature behavior is qualitatively the same as the one for the Schwarzschild-AdS black hole in the case with a sphere horizon ($k=1$).

As a result, the temperature behavior of the conformal anomaly corrected black holes with $k=-1$ is quite different from its counterpart
of the Schwarzschild-AdS black hole with $k=-1$. The latter behaves like the case with $\tilde{\alpha}/l^2<1/24$: its temperature
starts from zero at the minimal horizon $r_2=l/\sqrt{3}$ and increases monotonically, the black hole is always  local stable thermodynamically.

To see the  global stability of the black hole thermodynamics, we plot in Fig.~\ref{F_{k=-1,a=0.1/0.042}} the free energy
of the black hole with $\tilde{\alpha}/l^2=0.1$ (left plot) and $\tilde{\alpha}/l^2=1/24$ (right plot). We see that in both cases,
the free energy is always negative for larger black holes, while it is always positive for smaller black holes. At some horizon radius
the free energy changes its sign, where the Hawking-Page phase transition happens. Note that here the qualitative behavior is  same
for different choices of the integration constant $A_0$.  In Fig.~\ref{F_{k=-1,a=0.02}} we plot the free energy for the case with $\tilde{\alpha}/l^2=0.02$. Note that in this case there exists a minimal horizon radius $r_2 \approx 0.54$. We see that the free energy is always negative and the Hawking-Page transition is absent,  which means that the black hole with $k=-1$ and $\tilde{\alpha}/l^2=0.02$ is not only local stable, but also global stable thermodynamically, and the black hole phase  is always dominated in the dual conformal field theory side. It is worth mentioning here that the appearance of the Hawking-Page transition for the trace anomaly corrected AdS black holes with $k=-1$ is a new feature, to the best of our knowledge, such a feature for AdS black hole with $k=-1$ has not been reported previously in the literature.

\begin{figure}[h]
\centering
\includegraphics[scale=0.82]{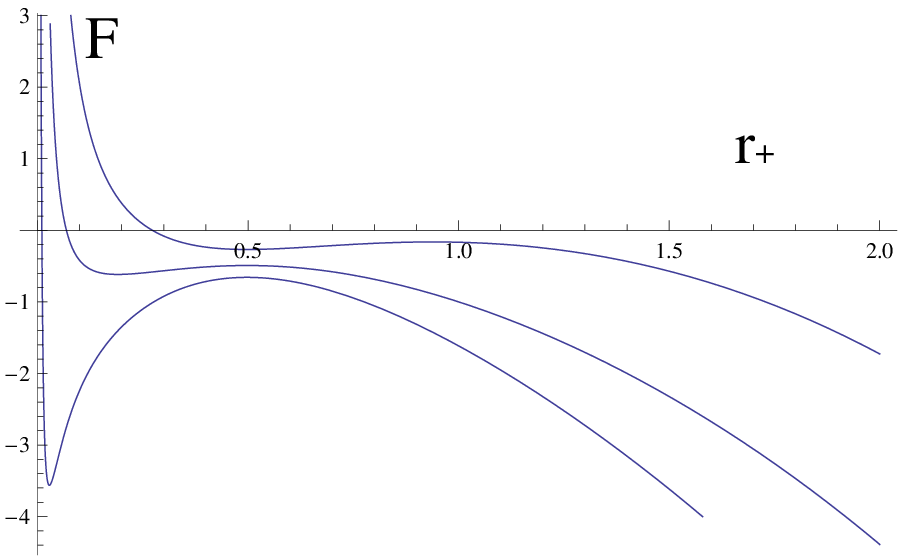}
\includegraphics[scale=0.82]{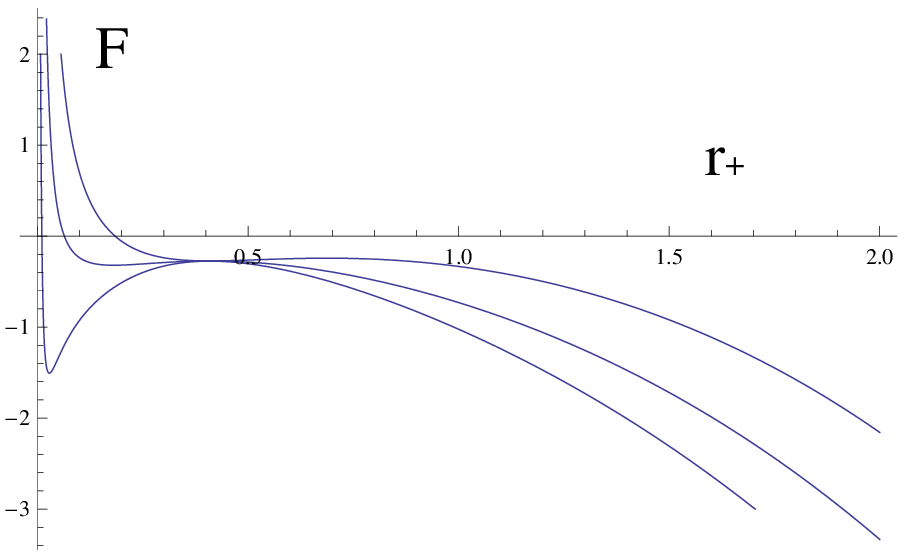}
\caption{\label{F_{k=-1,a=0.1/0.042}}The free energy of the black holes with $k=-1$ and $\tilde{\alpha}/l^2=0.1$ (left) and $\tilde{\alpha}/l^2=1/24$ (right). In both plots the three curves from up to down in the left side correspond to the cases with $A_0/l^2=100$, $0.5$ and $0.01$, respectively. Here the  horizon radius is scaled by $1/l$, while the free energy is scaled by $l/G$.}
\end{figure}

\begin{figure}[h]
\centering
\includegraphics[scale=0.82]{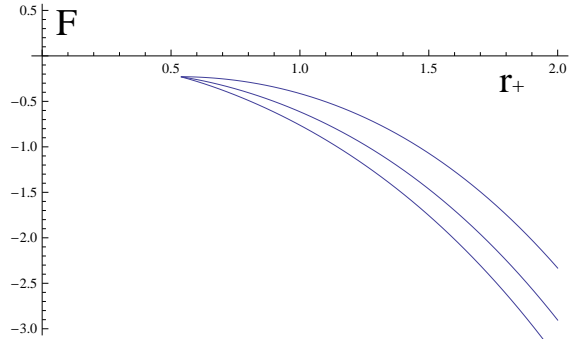}
\caption{\label{F_{k=-1,a=0.02}}The free energy of the black holes with $k=-1$ and $\tilde{\alpha}/l^2=0.02$. In this case, the minimal horizon $r_2/l \approx 0.54$. The three curves from up to down  correspond to the cases with $A_0/l^2=100$, $0.5$ and $0.01$, respectively. Here the  horizon radius is scaled by $1/l$, while the free energy is scaled by $l/G$.}
\end{figure}

 Now let us mention another interesting property for the black holes with $k=-1$. We can see from (\ref{eq13}) that even if $M=0$, the solution (\ref{eq10}) still describes a black hole with horizon
\begin{equation}
\label{eq25}
r_+^2= \frac{l^2}{2}\left(1+\sqrt{1+\frac{8\tilde {\alpha}}{l^2}}\right).
\end{equation}
This is just the so-called massless black hole in AdS space~\cite{Birm}.
In addition,
when $\tilde{\alpha}/l^2 < 1/24$, for the black hole
with minimal horizon $r_2$, its corresponding mass is
\begin{equation}
M_{\rm min}=-\frac{2l}{3\sqrt{6}G}\left(1-\frac{1}{2}\sqrt{1-\frac{24\tilde{\alpha}}{l^2}}\right) \sqrt{1+\sqrt{1-\frac{24\tilde{\alpha}}{l^2}}}.
\end{equation}
It is negative. The solution with $ M_{\rm min} \le M <0$ is called negative mass black hole~\cite{Birm}, while it is a naked singularity solution
if $M<M_{\rm min}$. If $\tilde {\alpha}=0$, one has $M_{\rm min}=-l/(3\sqrt{3}G)$.


\section{Conclusions and Discussions}
\label{sect4}
In this paper we have presented exact analytical black hole solutions with trace anomaly in AdS space. The black hole horizon can be
a positive, zero and negative constant curvature space, respectively.
We have studied thermodynamics of these black hole solutions and found that
the trace anomaly can qualitatively change the thermodynamical properties of these AdS black holes. These changes crucially depend on
the horizon structure of these black holes.

For the trace anomaly corrected AdS black hole with a positive curvature horizon, there exists a minimal horizon radius $2\sqrt{\tilde\alpha}$, determined by the degrees of freedom of the conformal field theory, which leads to the trace anomaly. The black hole
with the minimal horizon has a divergent Hawking temperature. These black hole have qualitatively similar thermodynamical properties as
those of the Schwarzschild-AdS black holes: There exists a minimal temperature, under which there does not exist any black hole solution; the black holes with small horizons are local unstable, while they become local stable for larger
horizons. But, depending on an unspecified integration constant $A_0$ in the black hole entropy [see (\ref{eq16})], the free energy of the
black holes can change its sign zero, one and two times when the black hole horizon varies, respectively. This seemingly implies that one has zero, one and two Hawking-Page phase transitions in this system.

For the trace anomaly corrected AdS black hole with a Ricci flat horizon, its thermodynamical behavior is completely the same as the Schwarzschild-AdS black hole with a Ricci flat horizon, although the two solutions are different. The trace anomaly corrected AdS black hole
is always local and global stable thermodynamically. The Hawking-Page transition is absent in this case.

For the trace anomaly corrected AdS black hole with a negative constant curvature horizon, its thermodynamical behavior depends on the
parameter $\tilde{\alpha}/l^2$: When $\tilde{\alpha}/l^2<1/24$, there exists a minimal horizon with vanishing Hawking temperature, the black hole with horizon larger than the minimal radius is always local and global stable thermodynamically, in this case, there is no Hawking-Page
transition. When $\tilde{\alpha}/l^2 \ge 1/24$, however, there is no the minimal horizon limit, the black holes with  smaller horizons are local unstable, while they are local stable for  larger horizons. The Hawking-Page phase transition always happen independent of the
integration constant $A_0$. This is a new feature for the trace anomaly corrected AdS black hole with a negative constant curvature horizon.
Such a feature has not been reported in the previous studies for the AdS black holes with negative constant curvature horizon in the
literature, to the best of our knowledge.

It should be mentioned here that we are able to present the analytical black hole solution in AdS space, only for the case with
the type A anomaly~\cite{DS}, namely the case with $\beta=0$ in (\ref{eq1}).  When $\beta \ne0$, one might have to solve the Einstein equations numerically with the assumption (ii). In addition, it is called for to make sure whether the solution with the ``+" branch in (\ref{eq10}) is dynamically unstable, like the case in Gauss-Bonnet gravity~\cite{Cai1}.

Here let us make some comments on the unspecified constant $A_0$ in the black hole entropy (\ref{eq16}). As we have already seen,
 the constant plays an important role in determining the global stability of the black hole thermodynamics. In principle, such a constant
should be determined by accounting the microscopic degrees of freedom of the black holes in quantum gravity, for example, in string theory~\cite{Sen}. In the present content of the note we are not able to specify the integration constant. But we still can make some discussions on this constant through some physical considerations. When $k=0$, one can take the constant $S_0=0$ as we did above, by
demanding the black hole entropy vanishes as its horizon goes to zero, because in this case, the logarithmic term in the black hole entropy (\ref{eq15}) is absent. When $k\ne 0$, as a measure of microscopic degrees of freedom, the black hole entropy has to be positive, which leads
us to have
\begin{equation}
\label{eq28}
A_0 \ge A  \exp [-A/(16\pi \tilde{\alpha}k)].
\end{equation}
As $k=1$, we note that there exists a minimal black hole horizon $2\sqrt{\tilde{\alpha}}$ with a divergent temperature. For such black holes, as argued by Holzhey and Wilczek in \cite{HW}, it is better to regard them as element particles. Take this viewpoint, one may think
the black hole with the minimal horizon should have zero entropy. In this case, we have $A_0=16\pi \tilde{\alpha}/e$.  Take this constant,
we plot in Fig.~\ref{F_{k=1,a=0.01,A0=0.1849}} the free energy for the black holes with $k=1$ in the cases of  $\tilde{\alpha}/l^2=0.01$, $0.04$ and $0.1$. In the cases of other $\tilde{\alpha}/l^2$, the behavior of the free energy is similar. We observe that in all cases,
the free energy is positive for smaller black holes, while it is negative for larger black holes, and the Hawking-Page phase transition
always happens.

\begin{figure}[h]
\centering
\includegraphics[scale=0.82]{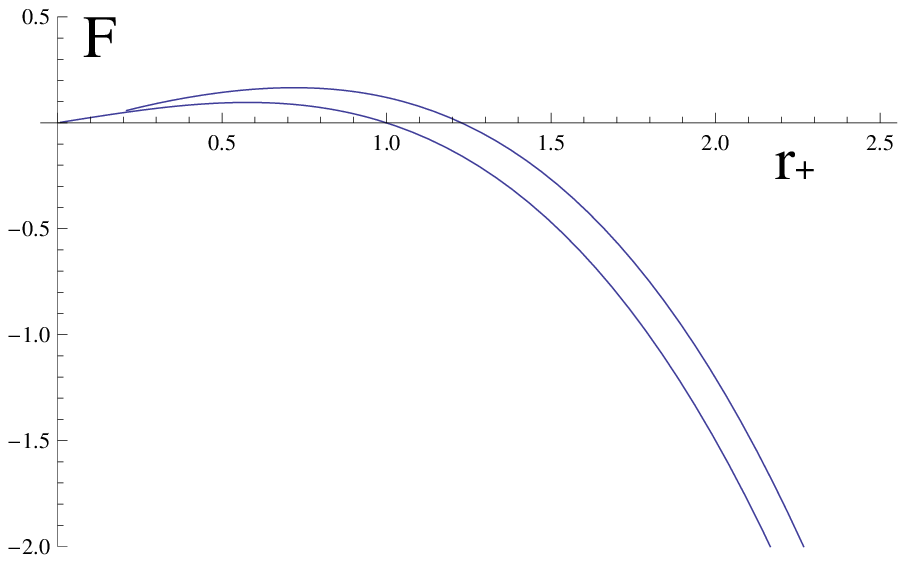}
\includegraphics[scale=0.82]{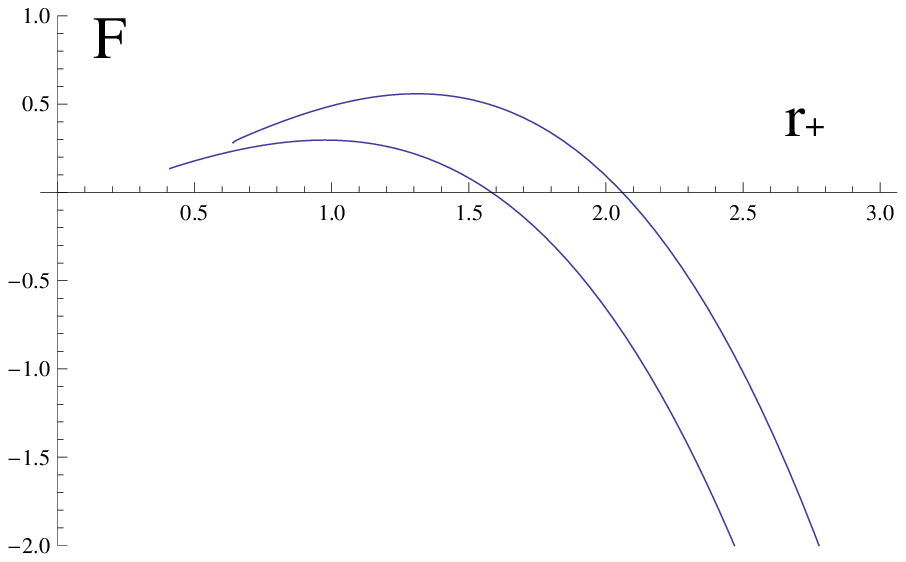}
 \caption{\label{F_{k=1,a=0.01,A0=0.1849}} The free energy of the black hole with $k=1$. Left: The upper curve corresponds to the case of $\tilde{\alpha}/l^2=0.01$ with $A_0/l^2 \approx 0.1849$, while the below one to the (Schwarzschild-AdS black hole) case of $\tilde{\alpha}/l^2=0$. Right: The upper curve corresponds to the case of $\tilde{\alpha}/l^2=0.1$ with $A_0/l^2 \approx 1.849$, while the below one to the  case of $\tilde{\alpha}/l^2=0.04$ with $A_0/l^2 \approx 0.7397$. Here the horizon radius is scaled by $1/l$, while the free energy is scaled by $G/l$. }
\end{figure}

As $k=-1$, we have no physical reason to choose a special black hole solution as the reference so that we can determine the constant $A_0$. Fortunately, in this case, as we have seen in Fig.~\ref{F_{k=-1,a=0.1/0.042}}, the different choices of $A_0$ do not qualitatively change the behavior of the free energy of the black holes with $k=-1$: the Hawking-Page transition always appears. It should be of great interest
to study the implication of the Hawking-Page transition in the dual conformal field theory side.

Finally we would like to stress the limitation and validness of the discussions in this note. Black hole thermodynamics is expected to be valid when a physical scale is much larger the Planck length $l_p$ ($\sim \sqrt{G}$). As we have seen, our parameter for the trace anomaly $\tilde \alpha \sim \alpha G$. Therefore to see the effect of the trace anomaly on the black hole thermodynamics, it has to be assumed that $\alpha \gg 1$. On the other hand, the thermodynamical feature of the Schwarzschild-AdS black hole is characterized by the ratio of horizon radius
to the AdS radius~\cite{HP,Witten}. Hence, in this paper we have introduced $\tilde{\alpha}/l^2$ and $r_+/l$ to characterize the back reaction strength of the trace anomaly and the size of the black hole, respectively. As a result, we believe that with the conditions $\tilde{\alpha}
\gg l_p^2$, $r_+\gg l_P$, and $l \gg l_p$, the thermodynamics of the trace anomaly corrected black holes makes sense in the semi-classical approximation. In addition, we point out here that the assumption (ii) is a strong restriction in order to find the exact analytical solution
of the equations (\ref{eq3}). Considering the full back reaction of the trace anomaly without the restriction, the horizon geometry might be
completely changed, and an interesting proposal is made in \cite{EM2}: At event
horizons quantum trace anomaly can have macroscopically large back reaction effects on the geometry, potentially removing the classical event horizon of black hole, replacing black hole horizon with a quantum phase boundary layer.  Such a picture has important consequences on the
black hole information loss paradox and currently observed dark energy in the universe.

\section*{Acknowledgements}

This work was supported in part by the National Natural Science Foundation of China (No.10821504, No.11035008, and  No.11375247).

\end{document}